\documentclass[twocolumn,prc,showpacs,preprintnumbers,superscriptaddress]{revtex4-1}

\usepackage{graphicx}
\usepackage{epsfig}
\usepackage{epstopdf}
\usepackage{dcolumn}
\usepackage{bm}
\usepackage{amsmath}
\usepackage{color}

\usepackage[utf8]{inputenc}
\usepackage[T1]{fontenc}
\usepackage{indentfirst}

\usepackage{enumitem}
\usepackage{xspace}
\usepackage{hyperref}

\newcommand{\Tiii}{t_0^\mathrm{III}}

\newcommand{\gras}[1]{\boldsymbol{#1}}

\usepackage{amssymb}
\usepackage{amsfonts}
\usepackage{mathrsfs}
\usepackage{units}
\usepackage{multirow}
\usepackage{braket}

\usepackage{bm}
\usepackage{rotating}

\newcommand{\disregard}[1]{}

\newsavebox{\tmpstrikebox}
\newlength{\tmpstrikelen}

\usepackage{tikz}
\newcommand{\tikzcircle}[2][red,fill=red]{\tikz[baseline=-0.5ex]\draw[#1,radius=#2] (0,0) circle ;}%

\begin{document}

\title{Precision calculation of isospin-symmetry-breaking corrections to $\gras{T=1/2}$ mirror decays using multi-reference charge-dependent density functional theory and beyond}

\author{M. Konieczka}
\affiliation{Institute of Theoretical Physics, Faculty of Physics, University of Warsaw, ul. Pasteura 5,
PL-02-093 Warsaw, Poland}

\author{P. B\k{a}czyk}
\affiliation{Institute of Theoretical Physics, Faculty of Physics, University of Warsaw, ul. Pasteura 5,
PL-02-093 Warsaw, Poland}

\author{W. Satu{\l}a}
\affiliation{Institute of Theoretical Physics, Faculty of Physics, University of Warsaw, ul. Pasteura 5,
PL-02-093 Warsaw, Poland}

\date{\today}

\begin{abstract}
\begin{description}
\item[Background]
        Theoretical approaches based on energy density functionals (EDF) are gaining on popularity due to their broad 
range of applicability. One of their key features is the proper treatment of symmetries of the nuclear interaction. It is because of this that EDF-based methods may provide another independent verification of foundations of the Standard Model, i.e.\ the assumption that the hadronic structure of matter is indeed built upon three generations of quarks. However, such a study cannot be precisely performed without including very subtle isospin-symmetry-breaking (ISB) terms both in the long- and short-range part of the nuclear interaction. Only very recently has the latter part of ISB interaction been successfully adapted to EDF to describe the Nolen-Schiffer anomaly.

\item[Purpose]
The aim of the paper is to study the impact of the short-range ISB terms on isospin impurities in the wave functions of $T=1/2$ mirrors ($\alpha_{\rm ISB}$) and the ISB corrections to their Fermi decays ($\delta_{\rm ISB}$). The consequent purpose is to eventually lead the calculation towards evaluation of the $V_{\rm ud}$ matrix element of the Cabbibo-Kobayashy-Maskawa (CKM) quark mixing matrix.

\item[Methods]
We use multi-reference density functional theory 
(MR-DFT)  that conserves angular momentum and properly treats isospin. Moreover, for the very first time, the functional includes both short- and long-range isospin-symmetry-breaking forces. Calculations are performed  using three different variants of the ISB interaction:
({\it i\/}) involving only the Coulomb force,   ({\it ii\/}) involving the Coulomb and 
leading-order (LO) contact isovector forces, and   ({\it iii\/}) involving the Coulomb and 
next-to-leading-order (NLO) contact isovector forces.
The evaluation of the $V_{\rm ud}$ matrix element requires a more subtle approach involving configuration mixing. For this reason, the calculation is performed with a DFT-rooted no core configuration interaction (DFT-NCCI) model including the above-mentioned ISB terms as well.

\item[Results]
We compute isospin impurities and ISB corrections in 
$T=1/2$ mirrors ranging from $A=11$ to 47 with MR-DFT formalism. 
Next, we focus on the best measured $A=19$, 21, 35, and 37 mirror pairs, calculate 
the ISB corrections to their Fermi decays with a DFT-rooted NCCI method, and then extract the $V_{\rm ud}$ matrix element. The final result shows that $V_{\rm ud}=0.9736(16)$, which 
(central value) is in good agreement with the value assessed from the superallowed $0^+ \to 0^+$ Fermi transitions and is  only slightly above the value obtained using a state-of-the-art shell model. Last but not least, we demonstrate the stability of our calculation.

\item[Conclusions]
The isovector short-range interaction surprisingly strongly influences the isospin impurities and ISB corrections in the $T=1/2$ mirrors as compared to  the calculation in which Coulomb interaction is the only source of isospin-symmetry breaking. Moreover, the $V_{\rm ud}$ matrix element is sensitive to the short-range isovector terms in the interaction and can be successfully extracted within the DFT-rooted approach that includes configuration mixing.

\end{description}
\end{abstract}

\maketitle

\section{Introduction}\label{intro}

With high-precision experiments and theoretical modeling of atomic nuclei one can test fundamental equations governing the properties of subatomic matter. Of particular interest are processes used to search for possible signals of {\it new physics\/} beyond the Standard Model (SM), like the superallowed $0^+\rightarrow 0^+$ $\beta$-decays, see~\cite{(Har15)} and references quoted therein.  With small, of order of a percent, theoretical corrections 
accounting for radiative processes and isospin-symmetry breaking (ISB),  these pure Fermi (vector) decays allow verification of the conserved vector 
current (CVC) hypothesis with a very high precision. In turn, they provide the most precise values of the leading element, $V_{\rm{ud}}$, of the Cabbibo-Kobayashi-Maskawa (CKM) matrix.

The mixed Fermi-Gamow-Teller decays of $T=1/2$ mirror nuclei, which are a subject of this work, offer an alternative way for the SM tests~\cite{(Sev08),(Nav09a)} 
provided that another observable like the $\beta$-neutrino correlation  ($a$),  $\beta$-asymmetry ($A$), or neutrino-asymmetry ($B$) coefficient is measured with high accuracy. 
The SM  expressions for $a_{\rm SM}$, $A_{\rm SM}$, and $B_{\rm SM}$ can be found, for example, in the review~\cite{(Sev06)}. 
These coefficients depend on angular momenta of the participating nuclear states and a mixing ratio $\varrho$
of the Gamow-Teller ($M_{\rm GT}$) and Fermi 
($M_{\rm F}$) matrix elements:
\begin{equation}
\varrho=\frac{g_{\rm A} M_{\rm GT}}{g_{\rm V} M_{\rm F}},
\end{equation}
where $g_{\rm V/A}$ are vector and axial-vector elecroweak currents coupling constants, respectively.  
Precision measurements of $a_{\rm SM}$, $A_{\rm SM}$, or $B_{\rm SM}$ provide, therefore, empirical values  
of $\varrho$, which are instrumental for the $V_{\rm ud}$ calculation because they allow us to avoid using theoretical values of $\varrho$ which, 
in spite of a recent progress  in the {\it ab initio\/} GT-decay calculation~\cite{(Gys19)}, are not yet accurate enough to be directly used for that purpose. With the experimental  $\varrho$ the $V_{\rm ud}$ calculation depends upon precision
measurement of the partial lifetime and theoretically calculated radiative ($\delta'_{\rm R},\delta_{\rm NS}^{\rm V}, \Delta_{\rm R}^{\rm V}$) and many-body ISB ($\delta_\textrm{ISB}^{\rm V}$) corrections to the Fermi branch. The latter are defined together with $\delta_{\rm NS}^{\rm V}$ as a deviation from the Fermi matrix element in its isospin-symmetry limit $M_F^0$:

\begin{equation}
|M_F|^2 = |M_F^0|^2 (1+\delta_{\rm NS}^{\rm V}-\delta_{\rm ISB}^{\rm V})
\label{delta}
\end{equation}

Indeed, the reduced lifetime for an allowed semileptonic  $\beta$-decay of $T=1/2$ mirror nuclei 
can be written as~\cite{(Sev08),(Nav09a)}:
\begin{eqnarray}
\mathcal{F} t^{\rm mirror}  & \equiv &   f_{\rm V} t (1+\delta'_{\rm R})(1+\delta_{\rm NS}^{\rm V} - \delta_\textrm{ISB}^{\rm V} )  \nonumber \\  
                                           & = & \frac{K}{G_{\rm F}^2 V_{\rm ud}^2  C_{\rm V}^2  (1+\Delta_{\rm R}^{\rm V}) \Big{(}1+\frac{f_{\rm A}}{f_{\rm V}}\varrho^2\Big{)}},
\label{ftmirror1}
\end{eqnarray}
where $K/(\hbar c)^6 = 2\pi^3 \hbar \ln 2 /(m_{\rm e} c^2)^5 =
8120.2787(11)\times 10^{-10}$\,GeV$^{-4}$s  is
a universal constant, $G_{\rm F}$ is the Fermi-decay coupling constant $G_{\rm F}/(\hbar c)^3 = 1.16637(1)\times 
10^{-5}$\,GeV$^{-2}$, and $f_{\rm V/A}$ denotes the phase space factors.  Hence, similar to the superallowed $0^+ \to 0^+$ decays, the quality of the test depends on the accuracy of empirical data and the quality of the theoretical models used to compute the corrections, in 
particular the many-body $\delta_\textrm{ISB}^{\rm V}$ corrections which are a subject of this work. 
Current precision of $T=1/2$ mirror decay experiments is achieved only for a handful of isotopes, which is still not enough for stringent 
testing of the SM. However, fast progress in  $\beta$-decay correlation techniques 
opens up new opportunities and keeps the field vibrant; see, for example, Ref.~\cite{(Fen18)} 
for the recent high-precision $\beta$-asymmetry measurement in $^{37}$K decay.

The goal of this work is to study the impact of isovector effective contact interaction that is adjusted to account for the Nolen-Schiffer anomaly~\cite{(Nol69)} in nuclear masses on isospin impurities in the wave functions of $T=1/2$ mirrors, the isospin symmetry breaking (ISB) corrections to their Fermi decays, and the $V_{\rm ud}$ matrix element in $sd$-shell $T=1/2$ mirror nuclei. We use different 
variants of symmetry-restored density functional theory (DFT) which were 
successfully applied in the past to compute isospin impurities and ISB corrections to super-allowed $0^+\rightarrow 0^+$ decays, see~\cite{(Sat09),(Sat11)}. 
After a brief presentation of the methods in Sect.~\ref{Methods} we demonstrate that the class-III local force strongly affects the calculated isospin 
impurities, see Sect.~\ref{impurities}. This rather counterintuitive observation motivated us to undertake a detailed theoretical study of the ISB corrections to the Fermi branch of $T=1/2$ ground state decays. In this context, in the first place, we present isospin and angular-momentum projected multi-reference DFT calculations covering $T=1/2$ nuclei with $11 \leq A \leq 47$, see Sect.~\ref{impurities}. Next, in Sect.~\ref{Vud}, we focus 
on decays of $A$=19, 21, 35, and 37 mirror nuclei for which experimental data on correlation parameters 
are precise enough to allow for extraction of the $V_{\rm ud}$ matrix element. For these cases we perform the DFT-based No-Core-Configuration-Interaction calculations (DFT-NCCI), see~\cite{(Sat16d)} for details,
including theoretical uncertainty analysis, see Sect.~\ref{error}.  The paper is summarized in Sect.~\ref{conclusions}.

\section{Methods}\label{Methods}

The nuclear mean-field-based models are almost perfectly tailored to study the ISB effects.
The single-reference DFT (SR-DFT) treats Coulomb polarization properly, without involving an approximation of an inert core, and accounts for an interplay between short- and long-range forces in a self-consistent way. The spontaneous symmetry breaking (SSB) effects that accompany the SR-DFT solutions and introduce, in particular, spurious isospin impurities and angular-momentum non-conservation can be then taken care of by extending the framework beyond mean field to multi-reference level (MR-DFT) with the aid of isospin- and angular-momentum projection techniques~\cite{(Sat09),(Sat10),(Sat11)}. However, the nuclear energy density functionals (EDF) which are conventionally applied in the DFT-based calculations use Coulomb as the only source of ISB. Therefore they are incomplete in the context of the ISB studies and cannot fully describe ISB observables like Triplet (TDE) or Mirror Displacement Energies (MDE) of nuclear binding energies. The latter deficiency is known in the literature as the Nolen-Schiffer anomaly~\cite{(Nol69)}. 
There is a consensus that these deficiencies cannot be cured without introducing non-Coulombic sources of ISB 
as shown within the nuclear shell model (NSM), Hartree-Fock (HF) theory or {\it ab initio\/} calculations in Refs.~\cite{(Orm89),(Suz93),(Bro00b),(Zuk02),(Car15),(Pet15),(Kan17),(Roc18)} and references given therein.

Recently, we constructed a single-reference charge-dependent DFT (SR-CDDFT) that includes, apart of the Coulomb and isoscalar Skyrme interactions, the leading-order (LO) zero-range and next-to-leading-order (NLO) gradient interactions of class~II, which introduces charge-independence breaking (CIB) and class~III describing charge-symmetry-breaking (CSB) effects in the Henley and Miller classification~\cite{(Hen79),(Mil95)}. We subsequently demonstrated that the SR-CDDFT allows for very accurate treatment of MDEs and TDEs in a very broad range of masses already in LO~\cite{(Bac18)} and showed that the description can be further improved by adding NLO terms ~\cite{(Bac19)}. 
In Ref.~\cite{(Bac19)} we also provide arguments that the newly introduced ISB terms
model strong-force-related effects of CIB and CSB rather than the beyond-mean-field electromagnetic corrections.

The aim of this work is to extend the SR-CDDFT to MR-CDDFT and  perform a systematic study of the isospin impurities and ISB corrections to the beta decays in $T=1/2$ mirrors. In this case the ISB effects due to class~II or class~IV forces are negligible~\cite{(Car15),(Bac18),(Bac19)}. Therefore, the non-Coulombic ISB force can be approximated by the isovector effective interaction up to NLO in the effective theory expansion:
\begin{align}\hspace*{-0.5cm}
\hat{V}^{\rm{III}}(i,j)  = 
\Big{(} & \Tiii\, \delta\left(\gras{r}_{ij} \right)
+ \frac12 t_1^{\rm{III}} 
\left[ \delta\left( \gras{r}_{ij} \right) \bm{k}^2 + \bm{k}'^2 \delta\left(\gras{r}_{ij} \right) \right]  \notag\\
+ & t_2^{\rm{III}} 
\bm{k}' \delta\left(\gras{r}_{ij} \right) \bm{k} \Big{)} \Big{(} \hat{\tau}_3^{(i)}+\hat{\tau}_3^{(j)}\Big{)} 
\label{eq:Skyrme_classIII}
\end{align}
where 
$\bm{k}  =  \frac{1}{2i}\left(\bm{\nabla}_i-\bm{\nabla}_j\right)$ 
($\bm{k}' = -\frac{1}{2i}\left(\bm{\nabla}_i-\bm{\nabla}_j\right)$) are relative momentum operators acting to the right (left), respectively.


The essence of MR-DFT is to cure the spurious effects of SSB.  The procedure boils down  to a rediagonalization of the entire Hamiltonian  in a good-isospin and good angular-momentum basis generated by acting on the HF configuration $|\varphi\rangle$
with the standard  1D isospin  $\hat P^T_{T_z T_z}$  and 3D angular-momentum  $\hat P^I_{MK}$ projection operators: 
\begin{equation}\label{IMKT}
|\varphi; \, IMK; \, T T_z\rangle 
    = \frac{1}{\sqrt{N_{\varphi; IMK; TT_z}}}  \hat P^T_{T_z T_z} \hat P^I_{MK} |\varphi \rangle  .
\end{equation}
Due to overcompleteness of the set (\ref{IMKT}), 
the rediagonalization of the Hamiltonian is performed by solving the Hill-Wheeler-Griffin 
equation in the {\it collective space\/} $-$ a subspace spanned by the linearly independent {\it natural states\/}
$|\varphi; \, IM; T T_z\rangle^{(i)} $ accounting for the 
$K$-mixing, see Ref.~\cite{(Dob09d)} for  further details. The resulting eigenfunctions are:
\begin{equation}\label{IMstates}
|n; \, \varphi ; \, IM; \, T_z\rangle =  \sum_{i, T\geq |T_z|}
   b^{(n I; \varphi)}_{i T} |\varphi;\, IM; TT_z\rangle^{(i)},
\end{equation}
where $n$ enumerates  eigenstates in ascending order according to their energies. The quantum state~(\ref{IMstates}) is free from spurious  isospin mixing. It has to be noted, however, that it may not be equipped with the correlation coming from higher excitations. Such a lack can be compensated by applying configuration mixing in DFT-NCCI formalism.

The DFT-NCCI scheme proceeds as follows. One starts with computation
of relevant (multi)particle-(multi)hole deformed HF configurations ${\varphi_i}$. Next, with the aid of projection methods, one
computes a set of projected states {$|\varphi_i; \, IMK; \, T T_z\rangle$}, see Eq.~(\ref{IMKT}), which are subsequently
mixed to account
for $K$-mixing and physical isospin-mixing. At this stage one obtains a set of non-orthogonal states  {$|n; \, \varphi_i; \, IM; \, T_z\rangle$} of Eq.~(\ref{IMstates}), which are eventually mixed by solving the Hill-Wheeler-Griffin equation.  In the mixing we use the same Hamiltonian that was used to create the HF configurations. Further details concerning the 
DFT-NCCI scheme can be found in Ref.~\cite{(Sat16d)}.


All calculations presented below were done using the code {\sc hfodd}~\cite{(Sch17)} 
with the SV$_{\rm SO}$ Skyrme force, a variant of the SV EDF of Ref.~\cite{(Bei75)} with
the tensor terms included and the spin-orbit strength increased by a factor of 1.2 as proposed in Ref.~\cite{(Kon16)}. 
The code includes the local ISB EDF, in the LO and NLO variants, and allows for simultaneous 1D isospin 
and 3D angular-momentum projections, and is also equipped with the DFT-NCCI module. 
In the following we compare three variants of the calculations including different ISB forces: ({\it i\/}) involving 
only the Coulomb  force ($\hat V_{\rm C}$), ({\it ii\/}) involving the Coulomb and LO contact ISB forces 
($\hat V_{\rm C}+\hat V_{\rm LO}^{\rm III}$) , and  ({\it iii\/}) involving the Coulomb and NLO 
 local ISB forces ($\hat V_{\rm C}+\hat V_{\rm NLO}^{\rm III}$). These variants will be labeled by the acronyms C, LO, and NLO, respectively.

\section{Results}\label{results}

\subsection{The influence of zero-range isovector interaction on isospin impurities and
ISB corrections to Fermi $\beta$-decays}\label{impurities}

\begin{figure}[ht!]
\includegraphics[width=\columnwidth]{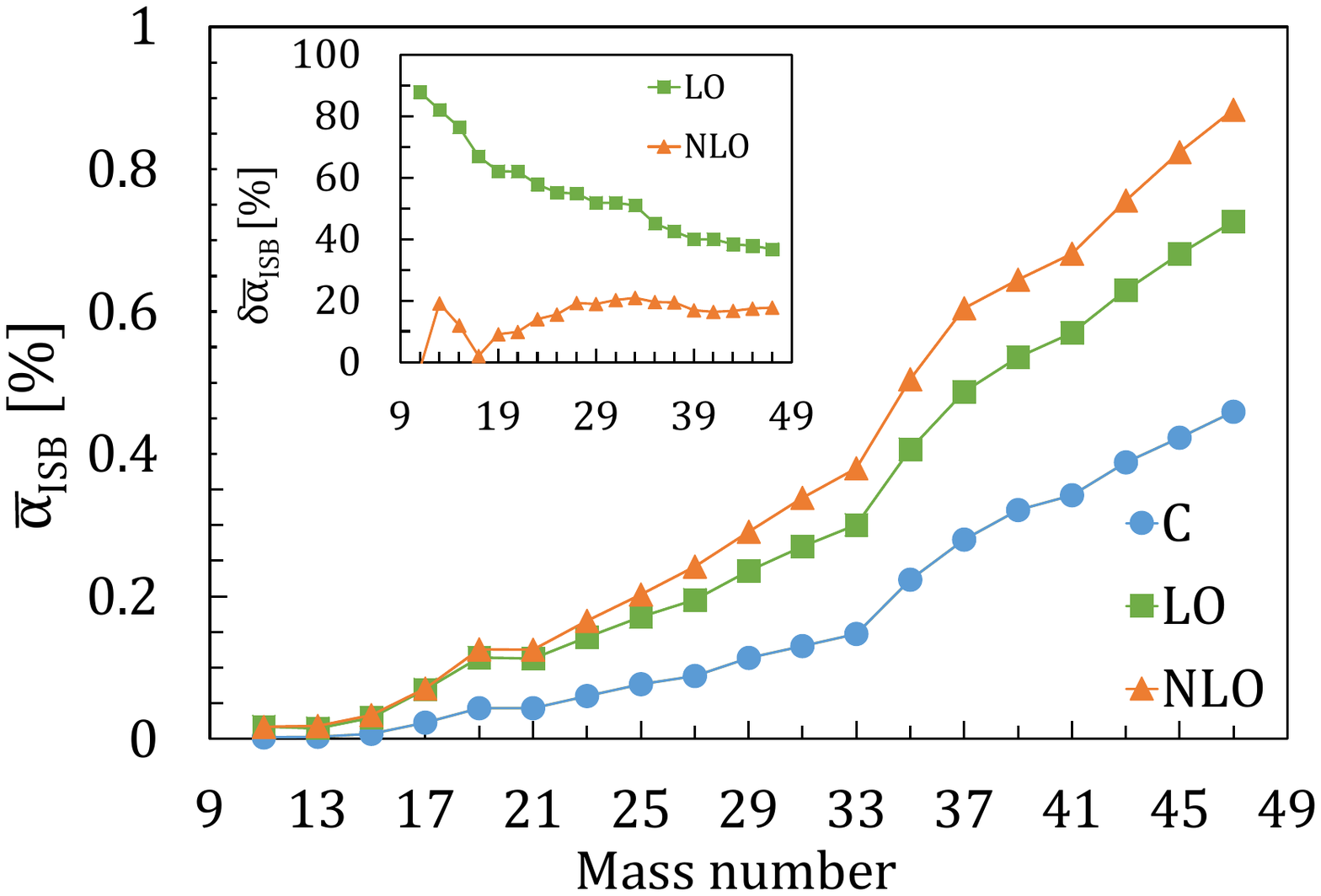}
\caption{(Color online) Arithmetic means of $\bar\alpha_{\rm C}$ (blue),  $\bar\alpha_{\rm LO}$ (green), and 
$\bar\alpha_{\rm NLO}$ (orange) over the ground-state values in $T_z=\pm 1/2$ mirror partners versus $A$. 
The insert shows relative differences $\delta\bar\alpha_{\rm LO} \equiv \frac{\bar\alpha_{\rm LO} - \bar\alpha_{\rm C}}{\bar\alpha_{\rm LO}}$ and $\delta\bar\alpha_{\rm NLO} \equiv \frac{\bar\alpha_{\rm NLO} - \bar\alpha_{\rm LO}}{\bar\alpha_{\rm NLO}}$.}
\label{fig:alphaC}
\end{figure}

The contribution to MDE in $T$=1/2 mirror nuclei due to the contact class-III interaction constitutes, on average, around 
7 to 8\% of the contribution coming from the Coulomb force,  as shown in Refs.~\cite{(Bac18),(Bac19)}. One would therefore 
naively expect that the class-III ISB force would also have a rather modest impact on the isospin impurity in 
the $n$-th state of spin $I$:  $\alpha_{\rm ISB}^{(n)} = 1- \sum_{i} |b^{(n I; \varphi)}_{i T=|T_z|}|^2$.  
Figure~\ref{fig:alphaC} shows arithmetic means
$$\bar{\alpha}_{\rm ISB}(A) = [{\alpha}_{\rm ISB} (A,T_z=1/2) + 
{\alpha}_{\rm ISB} (A,T_z=-1/2)]/2$$
in the ground states of $T_z=\pm 1/2$ for 
$11\leq A \leq 47$. The curves illustrate impurities obtained using C ($\alpha_{\rm C}$), LO 
($\alpha_{\rm LO}$), and NLO ($\alpha_{\rm NLO}$)  variants of the ISB interaction with parameters taken from~\cite{(Bac19)}.


It is surprising to see that the local class-III force strongly increases isospin mixing. The relative difference between $\alpha_{\rm LO}$ and $\alpha_{\rm C}$ gradually decreases  with $A$  (see insert in Fig.~\ref{fig:alphaC}) from 90\%
to circa 40\% (50\%) in the lower $fp$-shell nuclei for the LO (NLO) theory, respectively. Note also that the NLO theory brings a 
much smaller increase of $\alpha_{\rm ISB}$ as compared to the LO, which is expected for a converging effective 
theory. We have also verified (using isospin-projected theory) that a strong increase 
in $\alpha_{\rm ISB}$ due to class-III force takes place for density-dependent popular
Skyrme forces like SLy4~\cite{(Cha97)}.

\begin{figure}[ht!]
\includegraphics[width=\columnwidth]{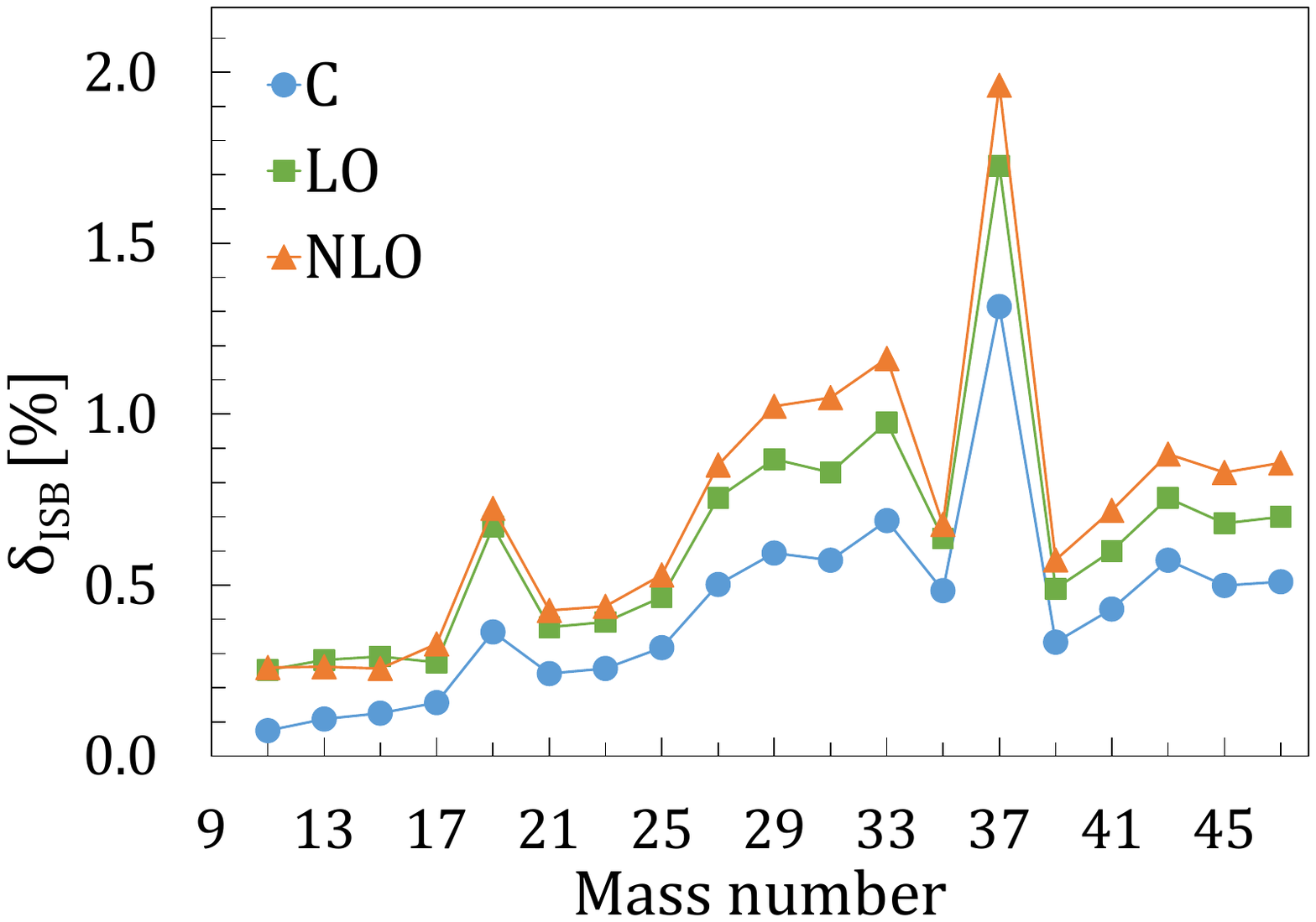}
\caption{(Color online)  ISB corrections to the Fermi branch of ground-state beta decay in $T=1/2$ mirror nuclei calculated 
using variants C (blue dots), LO (green squares), and NLO (orange triangles) of our MR-DFT model.}
\label{fig:deltaC}
\end{figure}

The additional isospin mixing introduced by ISB contact terms, see Eq.~(\ref{eq:Skyrme_classIII}), is expected to impact the ISB corrections to the Fermi branch in mirror $\beta$-decays. In order to assess the effect  quantitatively, we performed 
systematic calculation of $\delta_{\rm ISB}^{\rm V}$ in 
$11\leq A \leq 47$  using the SV$_{\rm SO}$ Skyrme force and three variants C, LO, and NLO of the ISB forces. Because the precision is of the utmost importance, we  refitted the class-III ISB interaction and adjusted its parameters to MDEs  in $11\leq A \leq 47$ calculated at the MR-DFT level.  
The fit gives $t^{\rm III}_0 = -6.3\pm 0.3$ MeV fm$^3$ for the SV$^{\rm LO}_{\rm SO}$ functional and  $t^{\rm III}_0 = 0 \pm 2$ MeV fm$^3$, $t^{\rm III}_1 = -2 \pm 2$ MeV fm$^5$, and 
$t^{\rm III}_2 = -4 \pm 1$ MeV fm$^5$  for the SV$^{\rm NLO}_{\rm SO}$ functional.
In the latter case we observed that the  $t^{\rm III}_0$ and $t^{\rm III}_1$ parameters are strongly correlated, which
increases their theoretical uncertainty and, in turn, the uncertainty on the calculated $\delta_{\rm ISB}^{\rm V}$. 
The results of the $\delta_{\rm ISB}^{\rm V}$ calculation are presented in Fig.~\ref{fig:deltaC}.  
As anticipated, an enhancement in $\alpha_{\rm ISB}$ implies strong 
enhancement  in  $\delta_{\rm ISB}^{\rm V}$, of the order of 70\% on average, caused by the LO term
 and further, albeit as expected much smaller, increase obtained in the NLO calculation.     


\subsection{Evaluation of the $V_{\rm ud}$ matrix element in DFT-NCCI calculation}\label{Vud}

The calculated $\delta_{\rm ISB}^{\rm V}$  versus the $A$ curve shows two irregularities for $A$=19 and $A$=37. Such irregularities indicate enhanced mixing among single-particle Nilsson orbitals and, indirectly, suggest that the MR-DFT calculations are not sufficient.
Similar problems were encountered  already in our seminal  MR-DFT calculation
of  $\delta_{\rm C}^{\rm V}$ for $0^+ \rightarrow 0^+$ superallowed Fermi decays
in  $A=38$ (and $A=18$) cases, see Ref.~\cite{(Sat11)}. The value of 
$\delta_{\rm C}^{\rm V}$ calculated in Ref.~\cite{(Sat11)} for $A=38$ was anomalously large due to accidental near-degeneracy and very strong mixing of the Nilsson orbitals originating from the $1s_{1/2}$ and $0d_{3/2}$ spherical sub-shells. Within the MR-DFT framework (involving projection from a single Slater determinant) such an anomalous case must be treated as an outlier and removed from further analysis of $V_{\rm ud}$, which was in fact done in Ref.~\cite{(Sat11)}. At the MR-DFT level of approximation there is no cure for such an effect. 
Hence, the large values of $\delta_{\rm ISB}$ for $A=37$ 
and, to a lesser extent, for $A=19$ mirror decays,  caused by the same $1s_{1/2} - 0d_{3/2}$ unphysical mixing,  should also be rejected from further analysis of $V_{\rm ud}$.  This, in turn, would limit the $V_{\rm ud}$ analysis within the MR-DFT to the two well-measured cases only, 
making the entire procedure statistically questionable.

With the development of DFT-NCCI~\cite{(Sat16d)}, however, 
we have at our disposal a new theoretical tool which allows us to control, at least to some extent, such an unwanted mixing. The model provides  rediagonalization of the entire Hamiltonian within the model space  that includes the interacting mean-field configurations. We, therefore, decided to analyze all four 
well measured $A$=19, 21, 35, and 37 mirror decays using this formalism. Moreover, in the present work DFT-NCCI calculations were limited to one-particle-one-hole ($p$-$h$)
configurations only. Such an approach was widely tested in our previous beta-decay calculation reported in~\cite{(Kon18)}.
The calculated ground-state (gs) and excited $p$-$h$ configurations in the four mirror nuclei are axially deformed, which implies that the number of participating $p$-$h$ configurations is very limited 
due to the $K$ quantum 
number conservation. In such a case, the configuration mixing, which proceeds through a spherically symmetric 
Hamiltonian, is effective only within the collective subspace built upon HF configurations 
of the same $K$. In $A$=19 and $A$=37 the gs configuration is built upon the $K$=1/2 Nilsson state with spin $I$=1/2 and $I$=3/2, respectively. Hence, the mixing is effective within three HF configurations having $\Delta K=0$. These configurations are built upon one of the 
three active $K$=1/2 Nilsson orbits $[220\, 1/2]$, $[200\, 1/2]$, and 
$[211\, 1/2]$  originating from the $d_{5/2}$,  $s_{1/2}$, and  
$d_{3/2}$ spherical sub-shells, respectively. 
In $A$=21 and $A$=35  the gs spin is $I$=$K$=3/2. The active model space then consists of only two HF configurations built upon either the 
$[211\, 3/2]$ or $[202\, 3/2]$ Nilsson orbits originating from $d_{5/2}$ and $d_{3/2}$ spherical subshells. 

\begin{figure}[ht!]
\includegraphics[width=\columnwidth]{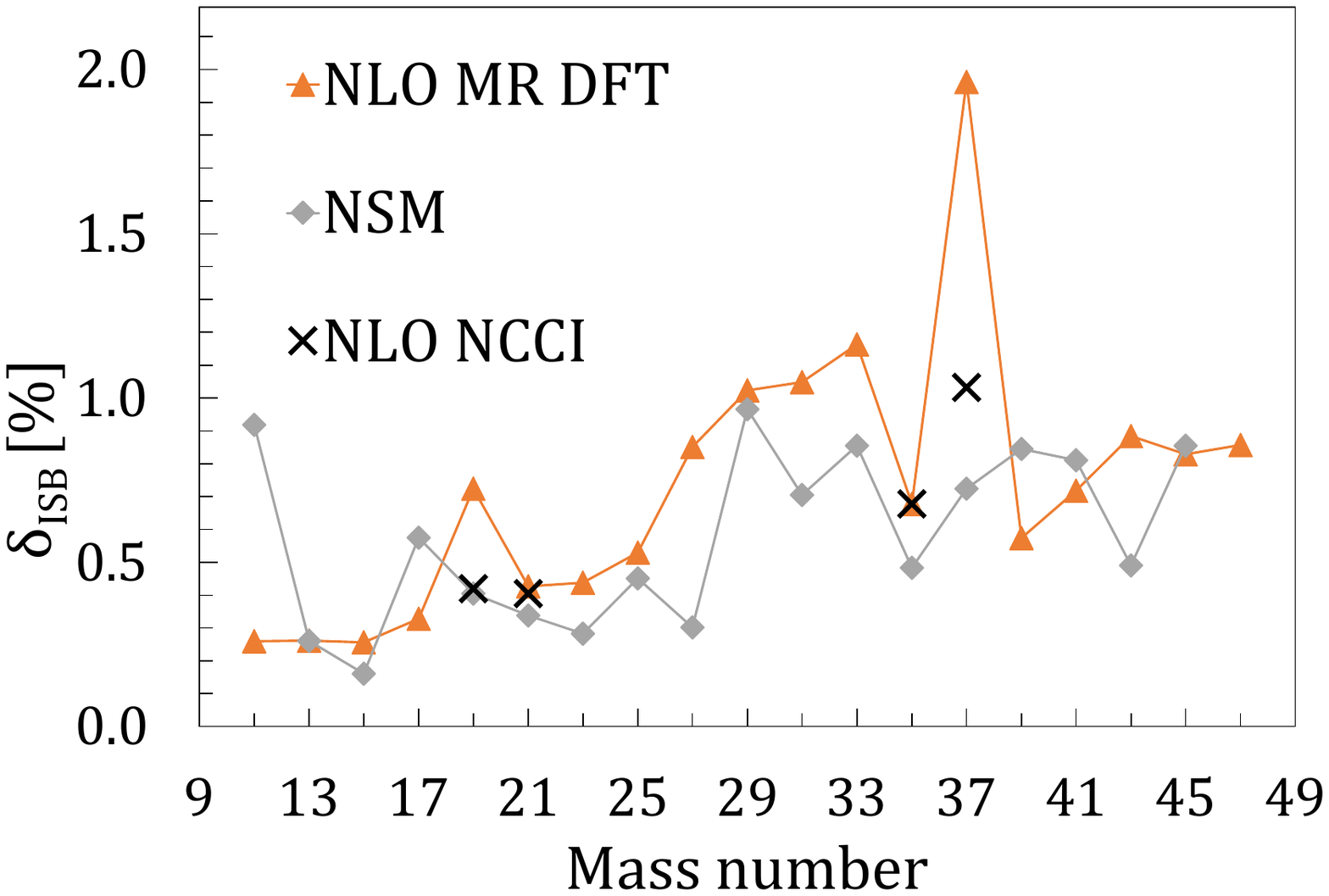}
\caption{(Color online)  ISB corrections to the Fermi branch of ground-state beta decay in $T=1/2$ mirror nuclei calculated 
using NLO (orange triangles) of our MR-DFT model in comparison with the 
nuclear shell model (NSM) results (grey diamonds) taken 
from Ref.~\cite{(Sev08)}. Black crosses mark the DFT-NCCI results for $A$=19, 21, 35, and 37 decays, see 
Tab.~\ref{tab:mixing}.}
\label{fig:deltaC2}
\end{figure}

The large energy gap between the $[211\, 3/2]$ or $[202\, 3/2]$ Nilsson orbits decreases considerably the effect of configuration mixing on $\delta_{\rm ISB}^{\rm V}$  in $A$=21 and  $A$=35. In contrast, the effect is very strong in $A$=19 and $A$=37. For the $^{19}$Ne$\rightarrow$ $^{19}$F, in the NLO variant, the 
ISB correction $\delta_{\rm ISB}^{\rm V}$ drops from 0.738\% calculated for the single configuration representing
the ground state to 0.580\% after admixing the first excited configuration, and further to 0.430\% after admixing
the second excited configuration. For the $^{37}$K$\rightarrow$ $^{37}$Ar decay the ISB correction decreases from 1.833\% to 1.099\% and down to 1.042\%, respectively. Note that the 
configuration mixing in $A$=19 and 37 corrects, to a large extent, the irregular behavior of $\delta_{\rm ISB}^{\rm V}$  
versus $A$ obtained for these two cases in MR-DFT, see Fig.~\ref{fig:deltaC2}.


\begin{table}[tbh]
\caption{ISB corrections $\delta^{\rm V}_{\rm ISB}$ to the Fermi transitions in 
$A$=19, 21, 35, and 37  calculated  using  the NSM~\cite{(Sev08)} and the C, LO, and NLO variants of the DFT-NCCI model. The last three rows show the results for  
$\bar{\mathcal{F} t}_0$, $V_{\rm ud}$, and for the unitarity test  obtained 
by averaging over the results in $A$=19, 21, 35,  and 37.}
\label{tab:mixing}
\renewcommand{\arraystretch}{1.3}
\begin{center}
\begin{tabular}{lcccc}
\hline
\hspace{0.27in} A    &   NSM  &  C  &  LO  &   NLO   \\
\hspace{0.27in} 19  
     &    0.415(39)    &   0.231(30)  &  0.412(54)  &  0.430(56)  \\
$\delta^{\rm V}_{\rm ISB}$ \hspace{0.01in} 21  
    &    0.348(27)    &   0.251(33)  &  0.394(50)  &  0.415(54)  \\
\hspace{0.27in} 35  
      &    0.493(46)    &   0.474(62)  &  0.647(84)  &  0.688(89)  \\
\hspace{0.27in} 37  
      &    0.734(61)     &   0.714(93)  &  0.97(13)  &  1.04(14)  \\
\hline
$\bar{\mathcal{F} t}_0$ 
     & \bf{6162(15)} & \bf{6166(18)} &  \bf{6156(18)} & \bf{6152(21)} \\
$V_{\rm ud}$  &  0.9727(14) &  0.9725(14)  &  0.9732(14)  &  0.9736(16) \\
unitarity  &  0.9967(31)  &  0.9961(31)  &  0.9976(31)  &  0.9983(35) \\
\hline
\end{tabular}
\end{center}
\end{table}

\begin{figure}[ht!]
\includegraphics[width=\columnwidth]{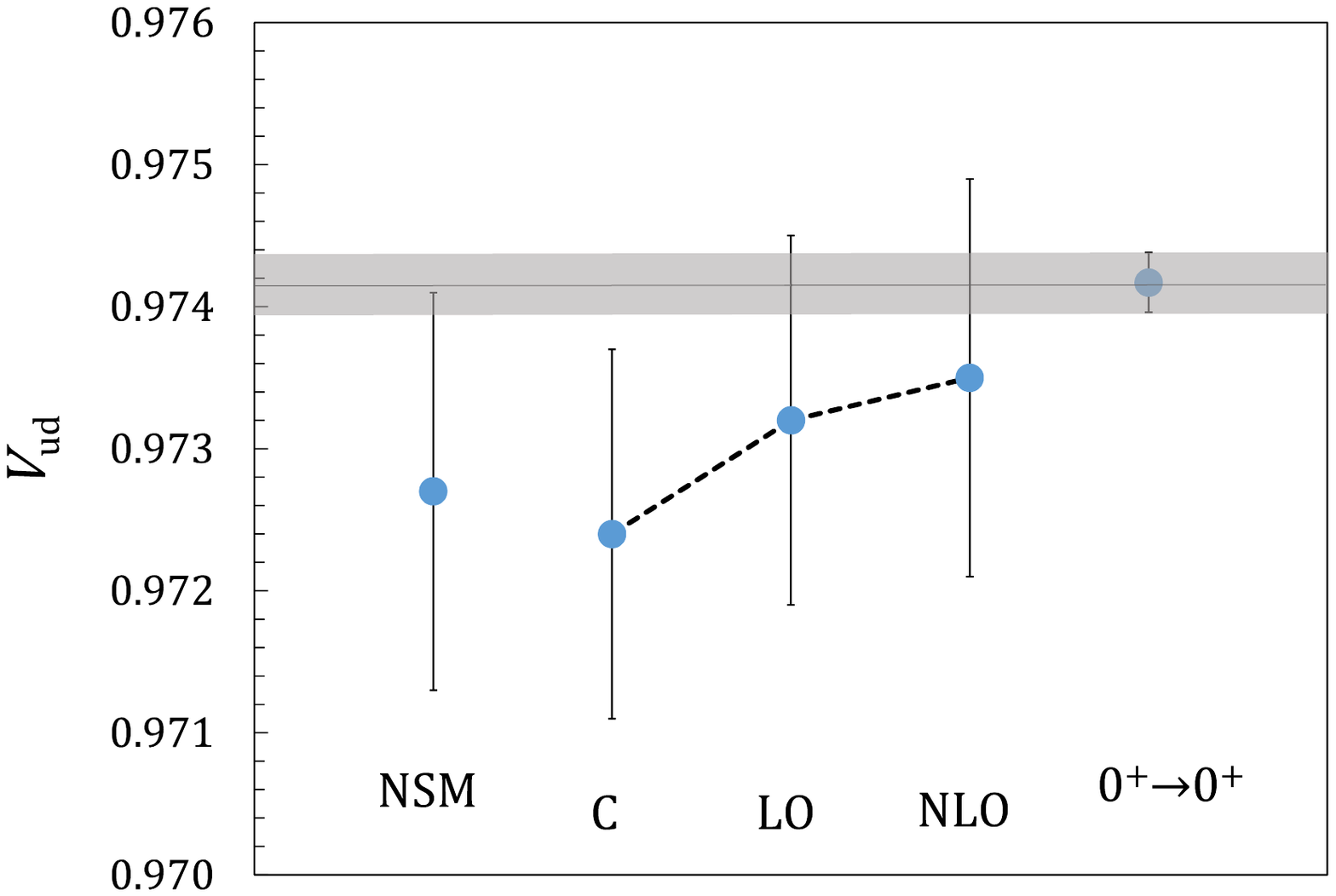}
\caption{(Color online) $V_{\rm ud}$ matrix element calculated from the  $T=1/2$, $A=$19, 21, 35, and 37 
mirror decays by means of the NSM~\cite{(Sev08)} and the
three variants C, LO, and NLO of the DFT-NCCI model.  Right point represents 
the  $V_{\rm ud}$  obtained from superallowed Fermi beta decays taken from Ref.~\cite{(Har15)}.} 
\label{fig:VUD}
\end{figure}

Table~\ref{tab:mixing} summarizes the results of DFT-NCCI calculations. It contains the results for the calculated 
values of  $\delta_{\rm ISB}^{\rm V}$, the average values of nucleus-independent reduced-lifetime $\bar{\mathcal{F} t}_0$ defined,
for a single transition, as:  
\begin{equation}
\mathcal{F}t_0  \equiv  \mathcal{F} t^{\rm mirror}\Big{(}1+\frac{f_{\rm A}}{f_{\rm V}}\varrho^2\Big{)}, 
\label{ftmirror2}
\end{equation}
the extracted values of $V_{\rm ud}$, and the result of the unitarity test.
Note that the DFT-NCCI theory 
for $V_{\rm ud}$ is convergent with respect to the addition of higher order ISB terms as depicted in Fig.~\ref{fig:VUD}, and that the final $V_{\rm ud}$ matrix element
\begin{equation}
V_{\rm ud}=0.9736\pm 0.0016 \nonumber
\end{equation}
lies within $\frac12 \sigma$ of the value assessed from superallowed $0^+ \to 0^+$ Fermi transitions, which is $V_{\rm ud}=0.97417\pm0.00021$~\cite{(Har15)}. In the calculations of $\bar{\mathcal{F} t}_0$ and $V_{\rm ud}$ we used 
the radiative corrections and phase-space factors taken  from Ref.~\cite{(Sev08)}.
The experimental data were taken from Ref.~\cite{(Tri12),(Bro14)} in the case of $A$=19 in which we have taken an error-weighted sum of a half-life time reported from these independent experiments; the same for $A=21$~\cite{(Gri15),(Shi18)}, $A=35$~\cite{(Nav09a)}, and for $A=37$~\cite{(Fen18)}.


\subsection{Theoretical uncertainty analysis}\label{error}

Let us finally comment on theoretical uncertainties.  The overall uncertainty imposed on the calculated ISB comes from 
three major sources: ({\it i\/}) from the cut-off on a harmonic oscillator basis, ({\it ii\/}) from the 
uncertainties of the class-III LECs, and finally ({\it iii\/}) from the configuration mixing. The uncertainties associated with the first two sources can be reliably estimated and do not exceed $\sim$5\%. The uncertainty associated with configuration mixing can be evaluated  
only {\it a posteriori\/}, after performing configuration-interaction calculations in the larger 
configuration space.

\begin{table*}[t!]
\centering
\caption{(Color online)  
Configurations used in the DFT-NCCI calculations for $A$=21 mirrors. Full dots denote pairwise occupied Nilsson states.  Up (down) arrows denote singly occupied 
Nilsson states with positive (negative) $K$ quantum numbers, respectively. 
Note that excitation of a pair to the $|202\, 5/2\rangle$ Nilsson level leads to 
oblate shape.}
\label{fig:A21CONF}
\begin{tabular}{|c|>{\raggedleft\arraybackslash}p{0.45cm}>{\raggedright\arraybackslash}p{0.45cm}|>{\raggedleft\arraybackslash}p{0.45cm}>{\raggedright\arraybackslash}p{0.45cm}|>{\raggedleft\arraybackslash}p{0.45cm}>{\raggedright\arraybackslash}p{0.45cm}|>{\raggedleft\arraybackslash}p{0.45cm}>{\raggedright\arraybackslash}p{0.45cm}|>{\raggedleft\arraybackslash}p{0.45cm}>{\raggedright\arraybackslash}p{0.45cm}|>{\raggedleft\arraybackslash}p{0.45cm}>{\raggedright\arraybackslash}p{0.45cm}|>{\raggedleft\arraybackslash}p{0.45cm}>{\raggedright\arraybackslash}p{0.45cm}|>{\raggedleft\arraybackslash}p{0.45cm}>{\raggedright\arraybackslash}p{0.45cm}|>{\raggedleft\arraybackslash}p{0.45cm}>{\raggedright\arraybackslash}p{0.45cm}||c|>{\raggedleft\arraybackslash}p{0.45cm}>{\raggedright\arraybackslash}p{0.45cm}|>{\raggedleft\arraybackslash}p{0.45cm}>{\raggedright\arraybackslash}p{0.45cm}|>{\raggedleft\arraybackslash}p{0.45cm}>{\raggedright\arraybackslash}p{0.45cm}|}\hline
group & \multicolumn{18}{c||}{PROLATE} & group & \multicolumn{6}{c|}{OBLATE} \\\hline   
config. & \multicolumn{2}{c|}{g.s.} & \multicolumn{8}{c|}{$\nu=1$} & \multicolumn{2}{c|}{\textit{nn/pp}}
& \multicolumn{2}{c|}{$\textit{np}$} & \multicolumn{2}{c|}{\textit{nn/pp}} & \multicolumn{2}{c||}{$\nu=3$} & config. & \multicolumn{2}{c|}{g.s.} & \multicolumn{2}{c|}{$\textit{nn/pp}$} & \multicolumn{2}{c|}{$\textit{np}$}\\\hline   
total $K$ & \multicolumn{2}{c|}{3/2} & \multicolumn{2}{c|}{1/2} & \multicolumn{2}{c|}{1/2} & \multicolumn{2}{c|}{1/2} & \multicolumn{2}{c|}{3/2} & \multicolumn{2}{c|}{3/2} & \multicolumn{2}{c|}{3/2} & \multicolumn{2}{c|}{3/2} 
& \multicolumn{2}{c||}{3/2} & total $K$ & \multicolumn{2}{c|}{3/2} & \multicolumn{2}{c|}{3/2} & \multicolumn{2}{c|}{3/2} \\\hline   
kernel & \multicolumn{2}{c|}{1} & \multicolumn{2}{c|}{2} & \multicolumn{2}{c|}{3} & \multicolumn{2}{c|}{4} & \multicolumn{2}{c|}{5} & \multicolumn{2}{c|}{6} & \multicolumn{2}{c|}{7} & \multicolumn{2}{c|}{8} & \multicolumn{2}{c||}{9} & kernel & \multicolumn{2}{c|}{10} & \multicolumn{2}{c|}{11} & \multicolumn{2}{c|}{12} \\\hline   
$|202\: 3/2\rangle$ & 
& &
& &
& &
& &
&
\hspace{0pt}{\color{red} $\bm\uparrow$}  &
& &
& &
& &
& 
& &
& &
& &
& \\  
$|211\: 1/2\rangle$ & 
& &
& &
& \hspace{0pt}{\color{red} $\bm\uparrow$} &
& &
& &
& &
& &
& &
\hspace{0pt}{\color{blue} $\bm\uparrow$} 
&
& &
& &
& &
& \\ 
$|200\: 1/2\rangle$ & 
& &
& &
& &
&\hspace{0pt}{\color{red} $\bm\uparrow$}  &
& &
& &
& &
\tikzcircle[blue, fill=blue]{0.7ex}  & &
\hspace{0pt}{\color{blue} $\bm\downarrow$} 
& 
& $|200\: 1/2\rangle$ & 
& &
& &
&\\ 
$|202\: 5/2\rangle$ & 
& &
& &
& &
& &
& &
& &
& &
& &
& 
& $|220\: 1/2\rangle$ & 
& &
& &
& \\ 
$|211\: 3/2\rangle$ & 
& \hspace{0pt}{\color{red} $\bm\uparrow$}  &
& \tikzcircle{0.7ex} &
&  &
&  &
&  &
\tikzcircle[blue, fill=blue]{0.7ex} & \hspace{0pt}{\color{red} $\bm\uparrow$} &
\hspace{0pt}{\color{blue} $\bm\uparrow$} & \tikzcircle{0.7ex} &
& \hspace{0pt}{\color{red} $\bm\uparrow$}  &
& \hspace{0pt}{\color{red} $\bm\uparrow$} 
& $|211\: 3/2\rangle$ & 
& \hspace{0pt}{\color{red} $\bm\uparrow$}  &
\tikzcircle[blue, fill=blue]{0.7ex}&  \hspace{0pt}{\color{red} $\bm\uparrow$} &
\hspace{0pt}{\color{blue} $\bm\uparrow$}   & \tikzcircle{0.7ex} \\ 
$|220\: 1/2\rangle$ & 
\tikzcircle[blue, fill=blue]{0.7ex} & \tikzcircle{0.7ex} &
\tikzcircle[blue, fill=blue]{0.7ex} & \hspace{0pt}{\color{red} $\bm\uparrow$} &
\tikzcircle[blue, fill=blue]{0.7ex} & \tikzcircle{0.7ex} &
\tikzcircle[blue, fill=blue]{0.7ex} & \tikzcircle{0.7ex} &
\tikzcircle[blue, fill=blue]{0.7ex} & \tikzcircle{0.7ex} &
                                                  & \tikzcircle{0.7ex} &
\hspace{0pt}{\color{blue} $\bm\uparrow$}& \hspace{0pt}{\color{red} $\bm\downarrow$} &
                                                  & \tikzcircle{0.7ex} &
                                                  & \tikzcircle{0.7ex} 
& $|202\: 5/2\rangle$ & 
\tikzcircle[blue, fill=blue]{0.7ex} & \tikzcircle{0.7ex} &
                                                  & \tikzcircle{0.7ex} &
\hspace{0pt}{\color{blue} $\bm\downarrow$} & \hspace{0pt}{\color{red} $\bm\uparrow$}  \\\hline  
\end{tabular}
\end{table*}

\begin{figure}[ht!]
\includegraphics[width=\columnwidth]{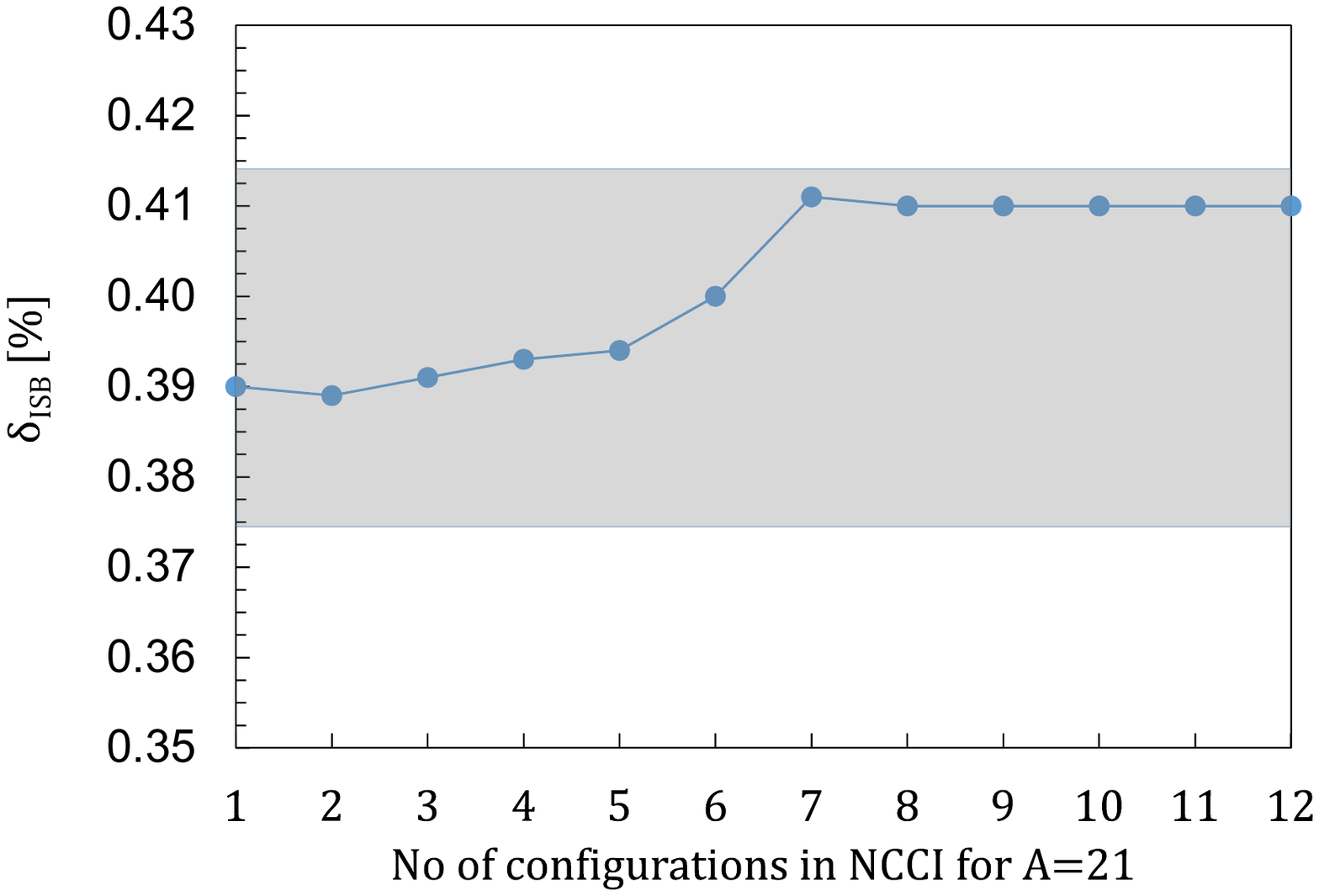}
\caption{(Color online) ISB correction to the Fermi transition $^{21}$Na$\to ^{21}$Ne with respect to an increase of a configuration space involving 1p$-$1h and 2p$-$2h pairing-type excitations. Configurations are added in the order listed in Table~\ref{fig:A21CONF}. Shaded area marks 5\% error bar superimposed on the DFT-NCCI
result calculated using 1p-1h configurations. The calculation was performed for the LO variant of the isovector interaction, see Eq.(~\ref{eq:Skyrme_classIII}) using the single-particle basis consisting of $N=12$ harmonic oscillators shells.} 
\label{fig:stability}
\end{figure}

In order to assess the uncertainty associated with configuration mixing we decided to perform  
configuration-interaction calculations for two representative cases $A=21$ and $A=37$.
In the case of $^{21}$Na$\to ^{21}$Ne decay we increased the model space to 12 
axially deformed configurations, which are depicted schematically in Tab.~\ref{fig:A21CONF}.
The result of the calculation is shown in Fig.~\ref{fig:stability}. The figure presents 
a relative change in the calculated ISB correction with respect to the value 
quoted in Tab.~\ref{tab:mixing}. As shown in the figure, admixture of 
seniority one 1p$-$1h excitations (configurations no 1-5) does not bring any relevant effect on $\delta_{\rm ISB}$. A perceptible increase can be noticed after admixing of $nn$-, $pp$-, and $np$- pairing-type 2p$-$2h excitations in the configuration space. The amount of the increase is around 5\%, 
which means that for this, and most likely, also for the $A$=35 case, our predictions can be considered 
as very stable with respect to the configuration mixing. 

\begin{figure}[ht!]
\includegraphics[width=\columnwidth]{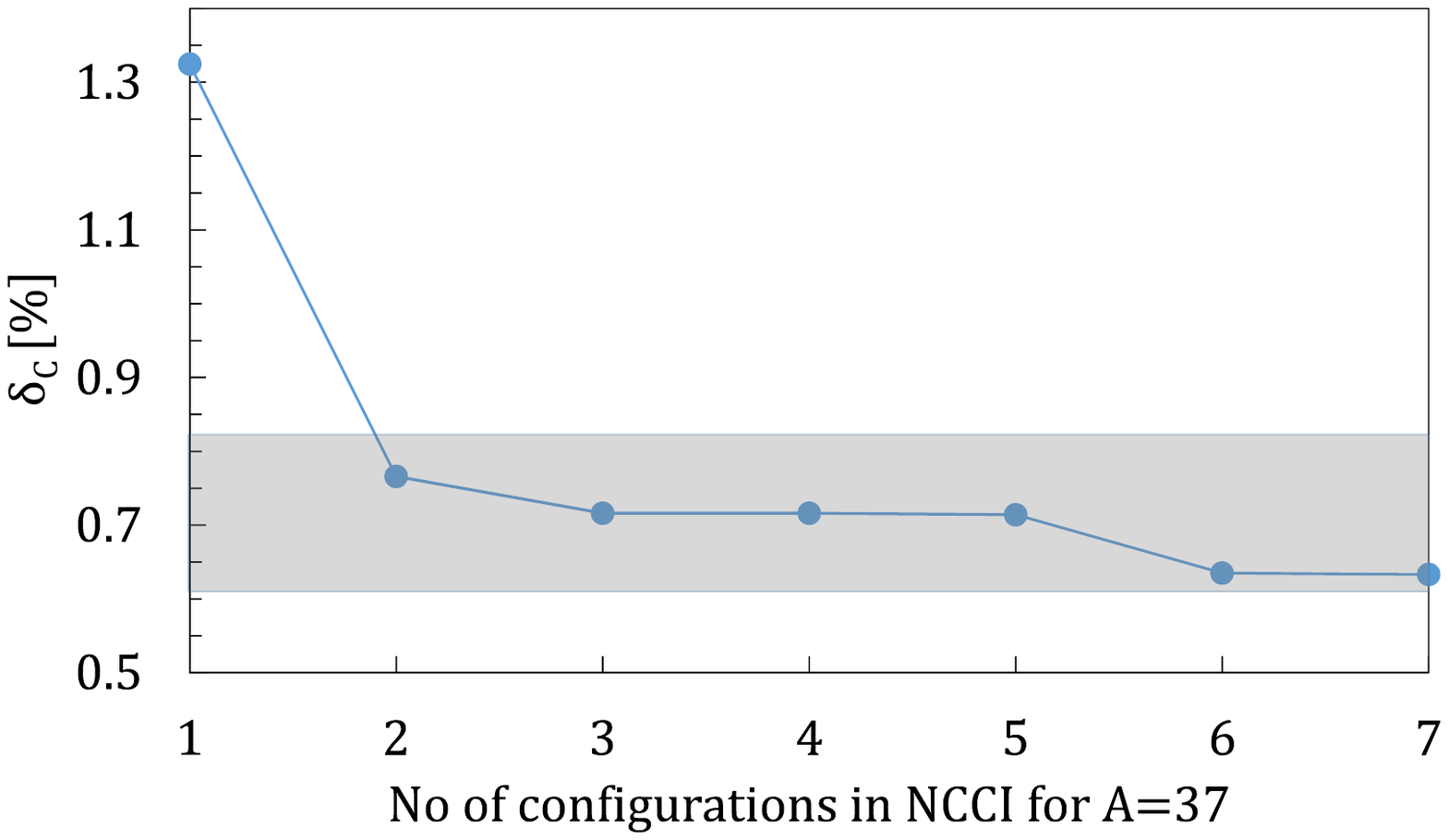}
\vspace{-0.3in}
\caption{(Color online) ISB correction to the Fermi transition $^{37}$K$\to ^{37}$Ar with respect to an increase of a configuration space involving 1p$-$1h and 2p$-$2h pairing-type excitations. Configurations are added in the order listed in Table~\ref{fig:A37CONF}. Shaded area marks 15\% error bar superimosed on the DFT-NCCI
result calculated using 1p-1h configurations. The calculation was performed  using the single-particle basis consisting of $N=12$ harmonic oscillators shells with the Coulomb as the only source of isospin-symmetry breaking.} 
\label{fig:stability2}
\end{figure}

We performed a similar calculation for the irregular and therefore most difficult case of  $^{37}$K$\to ^{37}$Ar transition.
In the analysis we included seven configurations depicted in Table~\ref{fig:A37CONF}.
The calculation indicates (see Fig.~\ref{fig:stability2}) that the total estimated error due to configuration 
mixing in this case is of the order of 15\%. It might be even slightly larger after 
including more $2p-2h$ excitations. However, the proximity of Nilsson orbitals provoking unstable HF solutions disables performance of such analysis. Nevertheless, at the moment, there is no strong motivation to conducting such a study. The theoretical error associated with the $\delta_{\rm ISB}$ calculation constitutes only a tiny fraction in the total error budget of  $|V_{\rm ud}|$, which is completely dominated by experimental uncertainties ~\cite{(Nav09a)}. 
In conclusion, we have imposed a 15\% error on our irregular $\delta_{\rm ISB}$ in $A=19$ and $A=37$ 
and a 5\% error in regular $A=21$ and $A=35$ cases due to configuration mixing. 

\begin{table}[h!]
\centering
\caption{(Color online)  
Configurations used in the DFT-NCCI calculations for $A$=37 mirrors. Full dots denote pairwise occupied Nilsson states.  Up (down) arrows denote singly occupied 
Nilsson states with positive (negative) $K$ quantum numbers, respectively.}
\label{fig:A37CONF}
\begin{tabular}{|c|>{\raggedleft\arraybackslash}p{0.35cm}>{\raggedright\arraybackslash}p{0.35cm}|>{\raggedleft\arraybackslash}p{0.35cm}>{\raggedright\arraybackslash}p{0.35cm}|>{\raggedleft\arraybackslash}p{0.35cm}>{\raggedright\arraybackslash}p{0.35cm}|>{\raggedleft\arraybackslash}p{0.35cm}>{\raggedright\arraybackslash}p{0.35cm}|>{\raggedleft\arraybackslash}p{0.35cm}>{\raggedright\arraybackslash}p{0.35cm}|>{\raggedleft\arraybackslash}p{0.35cm}>{\raggedright\arraybackslash}p{0.35cm}|>{\raggedleft\arraybackslash}p{0.35cm}>{\raggedright\arraybackslash}p{0.35cm}|}\hline
group & \multicolumn{14}{c|}{OBLATE} \\\hline   
config. & \multicolumn{2}{c|}{g.s.} & \multicolumn{8}{c|}{$\nu=1$} & \multicolumn{2}{c|}{\textit{nn/pp}}
& \multicolumn{2}{c|}{$\nu=3$} \\\hline   
total $K$ & \multicolumn{2}{c|}{1/2} & \multicolumn{2}{c|}{3/2} & \multicolumn{2}{c|}{1/2} & \multicolumn{2}{c|}{1/2} & \multicolumn{2}{c|}{3/2} & \multicolumn{2}{c|}{1/2} & \multicolumn{2}{c|}{1/2} \\\hline   
kernel & \multicolumn{2}{c|}{1} & \multicolumn{2}{c|}{2} & \multicolumn{2}{c|}{3} & \multicolumn{2}{c|}{4} & \multicolumn{2}{c|}{5} & \multicolumn{2}{c|}{6} & \multicolumn{2}{c|}{7} \\\hline   
$|211\: 1/2\rangle$ & 
 & \hspace{0pt}{\color{red} $\bm\uparrow$} &
 & \tikzcircle{0.7ex} &
 & \tikzcircle{0.7ex} &
 & \tikzcircle{0.7ex} &
 & \tikzcircle{0.7ex} &
\tikzcircle[blue, fill=blue]{0.7ex} & \hspace{0pt}{\color{red} $\bm\uparrow$} &
\hspace{0pt}{\color{blue} $\bm\downarrow$} & \hspace{0pt}{\color{red} $\bm\uparrow$}
\\  
$|202\: 3/2\rangle$ & 
\tikzcircle[blue, fill=blue]{0.7ex} & \tikzcircle{0.7ex} &
\tikzcircle[blue, fill=blue]{0.7ex} & \hspace{0pt}{\color{red} $\bm\uparrow$} &
\tikzcircle[blue, fill=blue]{0.7ex} & \tikzcircle{0.7ex} &
\tikzcircle[blue, fill=blue]{0.7ex} & \tikzcircle{0.7ex} &
\tikzcircle[blue, fill=blue]{0.7ex} & \tikzcircle{0.7ex} &
 & \tikzcircle{0.7ex} &
\tikzcircle[blue, fill=blue]{0.7ex} & \tikzcircle{0.7ex}
\\  
$|200\: 1/2\rangle$ & 
\tikzcircle[blue, fill=blue]{0.7ex} & \tikzcircle{0.7ex} &
\tikzcircle[blue, fill=blue]{0.7ex} & \tikzcircle{0.7ex} &
\tikzcircle[blue, fill=blue]{0.7ex} & \hspace{0pt}{\color{red} $\bm\uparrow$} &
\tikzcircle[blue, fill=blue]{0.7ex} & \tikzcircle{0.7ex} &
\tikzcircle[blue, fill=blue]{0.7ex} & \tikzcircle{0.7ex} &
\tikzcircle[blue, fill=blue]{0.7ex} & \tikzcircle{0.7ex} &
\tikzcircle[blue, fill=blue]{0.7ex} & \tikzcircle{0.7ex}
\\  
$|220\: 1/2\rangle$ & 
\tikzcircle[blue, fill=blue]{0.7ex} & \tikzcircle{0.7ex} &
\tikzcircle[blue, fill=blue]{0.7ex} & \tikzcircle{0.7ex} &
\tikzcircle[blue, fill=blue]{0.7ex} & \tikzcircle{0.7ex} &
\tikzcircle[blue, fill=blue]{0.7ex} & \hspace{0pt}{\color{red} $\bm\uparrow$} &
\tikzcircle[blue, fill=blue]{0.7ex} & \tikzcircle{0.7ex} &
\tikzcircle[blue, fill=blue]{0.7ex} & \tikzcircle{0.7ex} &
\hspace{0pt}{\color{blue} $\bm\uparrow$} & \tikzcircle{0.7ex} 
\\  
$|211\: 3/2\rangle$ & 
\tikzcircle[blue, fill=blue]{0.7ex} & \tikzcircle{0.7ex} &
\tikzcircle[blue, fill=blue]{0.7ex} & \tikzcircle{0.7ex} &
\tikzcircle[blue, fill=blue]{0.7ex} & \tikzcircle{0.7ex} &
\tikzcircle[blue, fill=blue]{0.7ex} & \tikzcircle{0.7ex} &
\tikzcircle[blue, fill=blue]{0.7ex} & \hspace{0pt}{\color{red} $\bm\uparrow$} &
\tikzcircle[blue, fill=blue]{0.7ex} & \tikzcircle{0.7ex} &
\tikzcircle[blue, fill=blue]{0.7ex} & \tikzcircle{0.7ex}
\\  
$|202\: 5/2\rangle$ & 
\tikzcircle[blue, fill=blue]{0.7ex} & \tikzcircle{0.7ex} &
\tikzcircle[blue, fill=blue]{0.7ex} & \tikzcircle{0.7ex} &
\tikzcircle[blue, fill=blue]{0.7ex} & \tikzcircle{0.7ex} &
\tikzcircle[blue, fill=blue]{0.7ex} & \tikzcircle{0.7ex} &
\tikzcircle[blue, fill=blue]{0.7ex} & \tikzcircle{0.7ex} &
\tikzcircle[blue, fill=blue]{0.7ex} & \tikzcircle{0.7ex} &
\tikzcircle[blue, fill=blue]{0.7ex} & \tikzcircle{0.7ex}
\\\hline
\end{tabular}
\end{table}

Let us finally mention that our results are obviously a subject to systematic error associated 
with  the form and parameters of the employed EDF. Our earlier calculations (see Ref.~\cite{(Sat10)}) as well as the random-phase approximation calculations by Liang {\it et al.\/}~\cite{(Lia09)} suggest that variations in EDF parameterizations should rather weakly influence 
the extracted $V_{\rm ud}$.  Detailed analysis of  such uncertainties, 
however, is very difficult and will not be performed here.  As already mentioned, 
the total error budget of  $|V_{\rm ud}|$ is, at present, dominated by experimental 
uncertainties ~\cite{(Nav09a)}.

\section{Conclusions}\label{conclusions}
 
In this paper, we have performed a systematic study of isospin impurities to the nuclear wave functions 
in $T$=1/2 mirror nuclei using MR-CDDFT, which includes, apart from the Coulomb interaction, the class-III ISB interaction adjusted to reproduce the Nolen-Schiffer anomaly in MDEs. 
We have investigated the impurities using three variants of the model including different ISB forces, namely: ({\it i\/}) involving  only the Coulomb  force, ({\it ii\/}) involving the Coulomb and LO contact ISB forces, and ({\it iii\/}) involving the Coulomb and local ISB forces up to NLO. We have demonstrated, for the first time, that the class-III interaction very strongly increases the isospin mixing, see Fig.~\ref{fig:alphaC}. Our results show that the NLO theory is convergent and brings a much smaller increase of $\alpha_{\rm ISB}$  as compared to the LO theory.

Next, we have presented a profound impact of class-III force on the isospin-symmetry-breaking corrections $\delta_{\rm ISB}^{\rm V}$ to the Fermi matrix elements of ground-state decays of $T$=1/2 mirror nuclei, which  constitute a theoretical input for the precision tests of the electroweak sector of the  SM.  In order to assess the effect quantitatively we have performed a systematic study of 
$\delta_{\rm ISB}^{\rm V}$ using MR-DFT with the three variants of the ISB force 
described above. As expected from the $\alpha_{\rm ISB}$ study, we have observed a strong systematic increase in  $\delta_{\rm ISB}^{\rm V}$ after including the LO class-III force and a further, albeit much smaller, increase within the NLO theory.  
 
The  $\delta_{\rm ISB}^{\rm V}$ calculated using MR-DFT shows  irregularities for $A$=19 and 37 cases, which are among the decays that are used for the SM test. Such irregularities usually indicate a mixing among the active
Nilsson orbitals, which can be taken care of by performing configuration-interaction calculations. In order to verify 
this conjecture and make our predictions more precise we performed the DFT-NCCI calculations of the ISB corrections in 
$A$=19, 21, 35, and 37 $T=1/2$ mirrors. Because these nuclei are axial we have limited the DFT-NCCI model space to
particle-hole deformed Nilsson configurations with $\Delta K$=0, with respect to the $K$ quantum number 
of the ground-state configuration. The DFT-NCCI results are shown in Tab.~\ref{tab:mixing}. 
The values of  $\delta_{\rm ISB}^{\rm V}$ calculated using the LO and NLO theories are systematically larger  than the results obtained using only the Coulomb interaction. They are also systematically larger than the corrections calculated using the NSM in Ref.~\cite{(Sev08)}. In turn, the extracted central value of $V_{\rm ud}$ matrix element 
is closer to the value obtained using  data on $0^+ \rightarrow 0^+$.  Our $|V_{\rm ud}|$=0.9736(16) was obtained with the error-weighted average over four mirror ($A$=19, 21, 35, and 37) transitions excluding the outlier $A$=29, a case measured  with lower accuracy as compared to other cases, see~\cite{(Nav09a)}. This value is considerably  larger than   $|V_{\rm ud}|$=0.9727(14), given in Ref.~\cite{(Fen18)} and 
above the value $|V_{\rm ud}|$=0.9730(14) of Ref.~\cite{(Gon19)}, which includes also the $A$=29 decay in the average. \\

\begin{acknowledgments}

This work was supported in part by the Polish National Science Centre
under Contract Nos.~2017/24/T/ST2/00160 and ~2018/31/B/ST2/02220.
We acknowledge CI\'S \'Swierk Computing Center, Poland, for the
allocation of computational resources.   

\end{acknowledgments}

\bibliography{VUD_MIRRORS,jacwit34}

\end{document}